\documentclass{article}
\usepackage{amssymb}
\usepackage{graphicx}
\usepackage{amsmath}

\input{tcilatex}

\begin{document}

\begin{center}
{\LARGE Extreme Test of Quantum Theory with Black Holes}

\bigskip\bigskip

\bigskip Antony Valentini

\bigskip

\bigskip

\textit{Perimeter Institute for Theoretical Physics,}

\textit{31 Caroline Street North, Waterloo, Ontario N2L 2Y5, Canada.}

(email: avalentini@perimeterinstitute.ca)
\end{center}

\bigskip

\bigskip

\bigskip

We propose an extreme test of quantum theory using astrophysical black holes
and entangled photons from atomic cascades. The identification of a cascade
emission close to a black-hole event horizon would allow us to observe
photons entangled with partners that have fallen behind the horizon. The
experiment involves testing the characteristic $\cos ^{2}\Theta $\
modulation of photon transmission through a pair of polarisers at relative
angle $\Theta $\ (Malus' law). For single photons, Malus' law is a
remarkable feature of quantum theory: it is equivalent to expectation
additivity for incompatible observables, and is generically violated for
hidden-variables theories with nonstandard probability distributions. An
experiment with entangled states straddling an event horizon is motivated by
the Hawking information loss puzzle, as well as on general grounds. In
principle, one could test the currently observed X-ray photons in iron lines
from black-hole accretion discs. However, only a small fraction ($\sim 0.6\%$%
) have cascade partners, and current X-ray polarimetry does not permit
successive measurements on a single X-ray photon. A realisable experiment
requires the identification of an appropriate cascade in a more convenient
frequency band.

\bigskip

\bigskip

\bigskip

\bigskip

\bigskip

\bigskip

\bigskip

\bigskip

\bigskip

\bigskip

\bigskip

\bigskip

\bigskip

\bigskip

\bigskip

\bigskip

\bigskip

\bigskip

\bigskip

\bigskip

\bigskip

\bigskip

\bigskip

\bigskip

\bigskip

\bigskip

\bigskip

\bigskip

\bigskip

\bigskip

\bigskip

\bigskip

\bigskip

\bigskip

\bigskip

\bigskip

\bigskip

\bigskip

\bigskip

\bigskip

\bigskip

\bigskip

\bigskip

\bigskip

\bigskip

\bigskip

\bigskip

\bigskip

\bigskip

\bigskip

\bigskip

Fundamental features of quantum theory, such as superposition and
entanglement, continue to be subjected to stringent tests. While quantum
theory shows no sign of breaking down, its interpretation remains as
controversial as ever, and it is important that the theory be probed in ever
more extreme conditions. The strong gravity region near macroscopic black
holes, in our Galaxy and in other galaxies, is being probed with increasing
precision, in particular by means of iron emission lines generated close to
the event horizon \cite{RN03}. This state of affairs affords an opportunity
to carry out a test of quantum theory in new and extreme conditions, if one
is able to identify an appropriate atomic emission close to a black-hole
horizon.

In a two-photon atomic cascade, an atom decays via an intermediate state,
generating a pair of photons with entangled polarisations. As is well known,
if the directions of the emitted momenta are appropriately constrained, the
polarisation state exhibits a strong and phase-coherent entanglement \cite%
{Asp02}. (Such states -- obtained from cascades in atomic calcium and
mercury -- were used in the classic early tests of Bell's inequality \cite%
{Asp02}.) By identifying such a cascade close to the horizon of a
macroscopic black hole, it would be possible to carry out an experiment
testing a key feature of quantum theory -- `expectation additivity for
incompatible observables', or equivalently Malus' law -- for single photons
that have the unusual property of being entangled with partners that have
fallen behind a black-hole event horizon. In appropriate circumstances,
there will be a significant probability that one of the cascade photons is
captured by the black hole while the other is detected (on Earth or on a
satellite).

Quantum theory predicts that single photons passing through a pair of
polarisers at relative angle $\Theta $ will be transmitted with a
probability that varies as $\cos ^{2}\Theta $. This modulation, or Malus'
law, is unremarkable in the classical theory of light; but for single
photons, it captures a remarkable property of quantum theory that was noted
very early by von Neumann and whose full significance was realised much
later by Bell.

Before considering practical details of the experiment, let us first explain
why Malus' law is remarkable in the case of single photons.

The polarisation of a single photon forms a two-state system. This may be
represented by a quantum observable $\hat{\sigma}=\mathbf{m}\cdot \mathbf{%
\hat{\sigma}}$, where $\mathbf{m}$ is a unit vector specifying a point on
the Bloch sphere and $\mathbf{\hat{\sigma}}$ is the Pauli spin operator. The
values $\sigma =\pm 1$ correspond respectively to polarisation parallel or
perpendicular to an axis $\mathbf{M}$ in physical space, where an angle $%
\theta $ on the Bloch sphere corresponds to a physical angle $\Theta =\theta
/2$. For an ensemble with density operator $\hat{\rho}$, the quantum
expectation value of $\mathbf{m}\cdot \mathbf{\hat{\sigma}}$ is given by the
Born probability rule as $\left\langle \mathbf{m}\cdot \mathbf{\hat{\sigma}}%
\right\rangle =\mathrm{Tr}\left( \hat{\rho}\mathbf{m}\cdot \mathbf{\hat{%
\sigma}}\right) =\mathbf{m}\cdot \mathbf{P}$, where the mean polarisation $%
\mathbf{P}=\langle \mathbf{\hat{\sigma}}\rangle $ (with norm $0\leq P\leq 1$%
) characterises the ensemble. With only two possible outcomes (transmission
or absorption), the fraction $p^{+}(\Theta )$ of photons transmitted through
a polariser set at angle $\Theta $ is fixed by the expectation value $%
\left\langle \mathbf{m}\cdot \mathbf{\hat{\sigma}}\right\rangle $ as%
\begin{equation}
p^{+}(\Theta )=\frac{1}{2}\left( 1+\left\langle \mathbf{m}\cdot \mathbf{\hat{%
\sigma}}\right\rangle \right) =\frac{1}{2}\left( 1+P\cos 2\Theta \right)
\label{Eqp}
\end{equation}%
For a fully-polarised beam ($P=1$), we have Malus' law $p^{+}(\Theta )=\cos
^{2}\Theta $.

The sinusoidal modulation in (\ref{Eqp}) reflects the dot-product structure $%
\mathbf{m}\cdot \mathbf{P}$ of the expectation value $\left\langle \mathbf{m}%
\cdot \mathbf{\hat{\sigma}}\right\rangle $. The origin of this structure may
be traced to a remarkable and fundamental feature of quantum theory, that
expectation values are additive for incompatible observables \cite{Sig}. If
the experiments $E_{1}$, $E_{2}$ respectively constitute quantum
measurements of non-commuting observables $\hat{\Omega}_{1}$, $\hat{\Omega}%
_{2}$, then $E_{1}$, $E_{2}$ will require macroscopically-distinct
experimental arrangements. Similarly, measurement of a linear combination $%
\alpha _{1}\hat{\Omega}_{1}+\alpha _{2}\hat{\Omega}_{2}$ (with $\alpha _{1}$%
, $\alpha _{2}$ real) will require a third experiment $E$,
macroscopically-distinct from $E_{1}$, $E_{2}$. Yet, over an ensemble with
density operator $\hat{\rho}$, the quantum expectation values -- obtained
from three different experimental arrangements -- will be related by $%
\langle \alpha _{1}\hat{\Omega}_{1}+\alpha _{2}\hat{\Omega}_{2}\rangle
=\alpha _{1}\langle \hat{\Omega}_{1}\rangle +\alpha _{2}\langle \hat{\Omega}%
_{2}\rangle $. And similarly for any linear combination of any number of
observables. Mathematically, this is a simple consequence of the linearity
of the Born rule, $\langle \hat{\Omega}\rangle =\mathrm{Tr}(\hat{\rho}\hat{%
\Omega})$ (for any $\hat{\Omega}$). But physically, it is a remarkable
relationship between results obtained from quite different experimental
arrangements. Indeed, expectation additivity is so powerful that the Born
rule may in fact be derived from it, as was first done by von Neumann \cite%
{vonN}. In the case at hand, for an arbitrary unit vector $\mathbf{m}%
=\sum_{i}c_{i}\mathbf{m}_{i}$, with $\mathbf{m}_{1}$, $\mathbf{m}_{2}$, $%
\mathbf{m}_{3}$ an orthonormal basis in Bloch space, expectation additivity
implies that $\left\langle \mathbf{m}\cdot \mathbf{\hat{\sigma}}%
\right\rangle =\sum_{i}c_{i}\left\langle \mathbf{m}_{i}\cdot \mathbf{\hat{%
\sigma}}\right\rangle $. Invariance of $\left\langle \mathbf{m}\cdot \mathbf{%
\hat{\sigma}}\right\rangle $ under a change of basis $\mathbf{m}%
_{i}\rightarrow \mathbf{m}_{i}^{\prime }$ then implies that $\left\langle 
\mathbf{m}\cdot \mathbf{\hat{\sigma}}\right\rangle =\mathbf{m}\cdot \mathbf{P%
}$ where $\mathbf{P}\equiv \sum_{i}\left\langle \mathbf{m}_{i}\cdot \mathbf{%
\hat{\sigma}}\right\rangle \mathbf{m}_{i}$ is a vector with norm $0\leq
P\leq 1$. Using expectation additivity again, we have $\mathbf{P}=\langle 
\mathbf{\hat{\sigma}}\rangle $. Thus (\ref{Eqp}) is equivalent to
expectation additivity \cite{Sig}.

That expectation additivity is remarkable (as well as powerful) becomes
clear when one considers, following Bell \cite{Bell66}, the individual
outcomes $\sigma \left( \mathbf{m}\right) =\pm 1$ for distinct,
linearly-related axes $\mathbf{m}$. For example, for (unit) $\mathbf{m}=c_{1}%
\mathbf{m}_{1}+c_{2}\mathbf{m}_{2}$, it is clearly impossible to satisfy the
condition $\sigma \left( \mathbf{m}\right) =c_{1}\sigma \left( \mathbf{m}%
_{1}\right) +c_{2}\sigma \left( \mathbf{m}_{2}\right) $ unless $c_{1}$ or $%
c_{2}$ vanishes. Thus if, in a given run of the experiment, the outcomes $%
\sigma \left( \mathbf{m}\right) $ along arbitrary axes $\mathbf{m}$ were
determined by some hidden variables $\lambda $, then for a subensemble of
experiments with fixed $\lambda $ the expectation values could not be
additive. It was argued by von Neumann \cite{vonN}, using assumptions that
included expectation additivity, that such `dispersion-free ensembles' were
mathematically inconsistent, and von Neumann concluded that the existence of
(any form of) hidden variables was incompatible with quantum theory.
However, by means of this simple example, Bell \cite{Bell66} showed that von
Neumann's assumption of expectation additivity was unreasonable for
(hypothetical) dispersion-free states.

Bell's example really illustrates a more general point, that expectation
additivity is -- from a hidden-variables perspective -- a peculiarity of a
particular distribution of variables $\lambda $. A (deterministic)
hidden-variables theory of a two-state system consists of some mapping $%
\sigma =\sigma \left( \mathbf{m},\lambda \right) $ (that determines the
outcome $\sigma =\pm 1$ of a measurement along an axis $\mathbf{m}$),
together with some assumed ensemble distribution $\rho _{\mathrm{QT}%
}(\lambda )$ of parameters $\lambda $, where $\rho _{\mathrm{QT}}(\lambda )$
is such that expectations $\int d\lambda \ \rho _{\mathrm{QT}}(\lambda
)\sigma \left( \mathbf{m},\lambda \right) $ agree with the quantum values $%
\left\langle \mathbf{m}\cdot \mathbf{\hat{\sigma}}\right\rangle $. An
arbitrary distribution $\rho (\lambda )\neq \rho _{\mathrm{QT}}(\lambda )$
of $\lambda $ will generally yield expectation values that disagree with
quantum theory. For an extreme case, where $\rho (\lambda )$ is concentrated
on just one value of $\lambda $, Bell's example shows that expectation
additivity fails. And for arbitrary ensembles with $\rho (\lambda )\neq \rho
_{\mathrm{QT}}(\lambda )$, expectation additivity will generally fail
because $\sigma \left( \mathbf{m},\lambda \right) \neq c_{1}\sigma \left( 
\mathbf{m}_{1},\lambda \right) +c_{2}\sigma \left( \mathbf{m}_{2},\lambda
\right) $ for every $\lambda $ (if $c_{1}c_{2}\neq 0$), and quantities that
are unequal for individual systems will generally remain unequal when
averaged over an arbitrary ensemble \cite{Sig}.

Such `non-quantum' or `non-equilibrium' distributions are not usually
considered, but their properties have been discussed [3, 6--12]. In the
pilot-wave theory of de Broglie and Bohm [13--15], for example, the
parameters $\lambda $ consist of configurations $X$ of particles or fields,
together with a guiding wave function $\Psi (X,t)$. Given a distribution $%
P=\left\vert \Psi \right\vert ^{2}$ of $X$ at some initial time, the
predictions of quantum theory follow. But in principle, one may consider
non-equilibrium distributions $P\neq \left\vert \Psi \right\vert ^{2}$,
yielding statistical results that deviate from quantum theory [6--9, 11, 12].

The key point here is that, in a hidden-variables theory, expectation
additivity for incompatible experiments is quite unnatural. It is, in fact,
a peculiarity of the special distribution $\rho _{\mathrm{QT}}(\lambda )$,
and is generally violated for $\rho (\lambda )\neq \rho _{\mathrm{QT}%
}(\lambda )$. The associated sinusoidal dependence in (\ref{Eqp}), for the
transmission of single photons through a polariser, is equally unnatural,
and is an exceptional property of the distribution $\rho _{\mathrm{QT}%
}(\lambda )$.

For single photons, then, Malus' law captures a peculiarly
quantum-mechanical phenomenon, and any test of Malus' law constitutes a test
of a fundamental feature of quantum theory, as fundamental and remarkable as
superposition or entanglement. Malus' law was tested for ordinary laboratory
photons by Papaliolios \cite{Pap}, for successive polarisation measurements
over short timescales $\sim 10^{-13}\mathrm{s}$, yielding agreement with $%
\cos ^{2}\Theta $ to within $1\%$. The test was motivated by a
hidden-variables theory of Bohm and Bub \cite{BB} (distinct from de
Broglie-Bohm theory) in which non-standard distributions of hidden variables
can exist for a short time immediately after a polarisation measurement.

Here, we propose a test of Malus' law for the more exotic case of photons
with entangled partners behind the event horizon of a black hole.

This may be motivated on general grounds, as a test of quantum theory in new
and extreme conditions. More specifically, it is not known how quantum
theory relates to gravitation; nor is it clear how to formulate quantum
field theory on background (classical) spacetimes with a nonstandard (or
non-globally-hyperbolic) causal structure, such as that widely believed to
be associated with the formation and evaporation of a black hole \cite{Wald}%
. The latter process arguably leads to pure quantum states evolving into
mixed states \cite{H76}, a result that conflicts with the usual rules of
quantum theory, and which leads to a failure of retrodictability of the
initial state from the final state. This Hawking `information loss' might be
avoided if nonstandard distributions of hidden variables were generated at
the exterior wing of entangled states straddling the horizon, since the
(non-quantum) statistics outside the hole could then contain more
information than that carried by an ordinary mixed quantum state \cite{AV04}%
. Such a process could occur, for example, in de Broglie-Bohm theory if (for
whatever reason) non-equilibrium degrees of freedom existed inside the black
hole: for if these interacted locally with the interior wing of an entangled
state, then the remote exterior wing would evolve away from quantum
equilibrium (the effect presumably taking place along some unknown spacelike
hypersurface) \cite{AV04}.

We now consider how the experiment might be realised.

It is believed that most galactic nuclei contain a supermassive black hole
(of mass in the range $\sim 10^{6}-10^{10}$ solar masses) accompanied by a
thin accretion disc. The strong gravity region close to the hole generates
X-rays, and the profiles of the X-ray emission lines may be used to probe
the details of the spacetime geometry at the location of the radiating
material. According to current models \cite{RN03}, a hot corona above the
disc irradiates the surface of the inner region with a continuum of X-rays,
causing the photo-ionisation of iron atoms at the surface of the disc. The
transition with the largest cross-section results in the ejection of a
K-shell ($n=1$) electron. An L-shell ($n=2$) electron can then fall into the
vacant K-shell (a vacancy transition $1s\rightarrow 2p$), with the emission
of a K$\alpha $ line X-ray photon at $6.4\ \mathrm{keV}$. This line has been
observed (by satellite) in detail for a number of active galaxies. The
intrinsically narrow (fluorescent) line is broadened and skewed, with an
extended red wing consistent with the effect of gravitational redshift on
photons emitted from very close to the horizon. It is usually assumed that
the disc does not extend all the way to the event horizon, but has an inner
edge close to or at the radius of marginal stability $r_{\mathrm{ms}}$ (the
radius of the last stable circular orbit). In most models, the observed
X-ray line emission cannot come from inside $r_{\mathrm{ms}}$. In some
systems, the extreme red wing of the iron line originates from radii that
are within a factor of 2 of the horizon radius. (For a review of iron lines
as probes of black-hole systems, see ref. \cite{RN03}.)

Now, if a cascade emission should generate entangled photon pairs at such
small radii, then a significant fraction of the photons reaching Earth will
have partners that were captured by the black hole: owing to atomic recoil,
the directions of the emitted photon momenta are not strongly correlated, so
that partners may be emitted over a wide range of angles. (For a black hole
of mass $M$ and specific angular momentum $a$, the event horizon radius is $%
r_{+}=M+\sqrt{M^{2}-a^{2}}$ while the photon capture cross section is $%
\sigma _{\mathrm{cap}}\approx 25\pi M^{2}$, using units where $G=c=1$ \cite%
{BHP}.) In principle, successive polarisation measurements of the received
photons could then be performed, realising the proposed extreme test. In
practice there are a number of difficulties, which do not, however, seem
insurmountable.

The test will be particularly interesting if the outgoing photon
polarisations are strongly entangled with the infalling photon
polarisations. In an atomic cascade, strong (phase-coherent) polarisation
correlations are obtained only if the photon momenta are appropriately
constrained. For example, in an EPR-Bell correlation experiment with a $%
0-1-0 $ cascade, if the detectors are placed collinearly on each side of the
source and have a negligible acceptance angle, then conservation of angular
momentum and parity enforce a polarisation state $\frac{1}{\sqrt{2}}\left( |%
\mathbf{\hat{x}}\rangle \otimes |\mathbf{\hat{x}}\rangle +|\mathbf{\hat{y}}%
\rangle \otimes |\mathbf{\hat{y}}\rangle \right) $ (where $|\mathbf{\hat{x}}%
\rangle $, $|\mathbf{\hat{y}}\rangle $ respectively denote states of
polarisation along the $x$-, $y$-axes, taking the source and detectors to
lie along the $z$-axis) \cite{Asp02}. Including non-antiparallel pairs of
photon momenta reduces the polarisation correlation coefficient, which is
generally equal to $F(\delta )\cos 2\phi $ where $\phi $ is the angle
between the polariser axes, $\delta $ is the half-angle subtended by the
detectors, and $\left\vert F(\delta )\right\vert $ ($\leq 1$) is a
decreasing function which depends on the cascade. In the experiments of
Aspect \textit{et al}. \cite{Asp02} with a $0-1-0$ cascade, $\delta =32%
{{}^\circ}%
$ and $F(\delta )=0.984$, showing that if the photon momenta are only
approximately antiparallel the correlation can still be very strong.

Approximately antiparallel momenta may be realised in our proposed
experiment (which of course requires polarisation measurements to be
performed only on the outgoing photons) by focussing attention on the
observed photons with the greatest redshift. As shown in detail by
Cunningham \cite{C75}, such photons have emission radii $r_{\mathrm{e}}$
closest to the horizon at $r_{+}$; and, as $r_{\mathrm{e}}\rightarrow r_{+}$%
, photons must be emitted parallel to the disc surface in order to avoid
being absorbed by the hole or the disc. These results hold for all
inclination angles of the disc with respect to the distant observer. (See
Figs. 1b and 5b of ref. \cite{C75}, for the case of a Kerr black hole with $%
a/M=0.998$. Note that we are considering only emissions taking place at or
near the surface of the disc.) The most redshifted photons detected on Earth
will have been emitted close to the horizon at $r_{\mathrm{e}}\approx r_{%
\mathrm{ms}}$ (no emission being expected between $r_{+}$ and $r_{\mathrm{ms}%
}$), in directions approximately parallel to the disc. (For example,
according to Fig. 5b of ref. \cite{C75}, an axial or face-on observer
receives photons from $r_{\mathrm{e}}\approx r_{\mathrm{ms}}\approx 1.2M$
that were emitted at an angle $\theta _{\mathrm{e}}\approx 80%
{{}^\circ}%
$ to the disc normal.) Some of these photons will have partners that
actually fell behind the horizon. For some of these pairs, the momenta will
be approximately oppositely directed at the point of emission, and the
polarisation entanglement will be strong (in the example of a $0-1-0$
cascade). The rest of the detected photons -- with partners that fell behind
the horizon but whose momenta were not approximately oppositely directed at
the point of emission, or with partners that did not fall behind the horizon
at all -- are expected to have standard polarisation probabilities, and
their presence will merely dilute the sought-for effect.

The entanglement between the outgoing and ingoing photons might be destroyed
by scattering along the line of sight, or by interaction with the infalling
material. However, entanglement has been shown to be surprisingly robust
against scattering \cite{AEW02}; though the momentum spread of the scattered
states does diminish the polarisation entanglement \cite{VB04}. If the line
of sight from emission to Earth is close to the plane of the accretion disc,
so that our view of the disc is close to edge-on, then scattering by dust in
the plane of the galaxy in question could be avoided by using an infrared
cascade. Alternatively, one might restrict attention to cases where the
accretion disc is viewed face-on. This would be an advantage because the
accretion discs in active galactic nuclei are surrounded by a co-aligned
dusty torus, so that a face-on view yields a clear line of sight to the
central engine \cite{Ant93}.

As for scattering behind the horizon, the experiment will be of particular
interest in cases where the ingoing photons reach deep into the interior
without encountering the infalling material. This can occur, depending on
the direction of emission of the photons. In the `standard model' of
geometrically thin accretion discs \cite{PT74}, the disc height $h(r)<<r$ at
every radius $r$. In the inner region, $h<<r\sim r_{\mathrm{ms}}\sim M$ (for
a near-extremal Kerr black hole, as observed in some galaxies \cite{RN03}),
so that the height of the disc is small compared to the critical impact
parameter $b_{\mathrm{cap}}=\sqrt{\sigma _{\mathrm{cap}}/\pi }\approx 5M$
for photon capture by the black hole. Thus, for geometrical reasons, photons
emitted from the surface of the disc (in appropriate directions) can avoid
the infalling material while nevertheless being captured, a process which
has an important effect on the evolution of the hole \cite{KT74}. (This
conclusion holds even if the disc is `thick', with $h(r)\lesssim r$ close to
the hole, as occurs in some models.) Again, in a real case photons will be
emitted in all directions, diluting the sought-for effect.

In some of the earlier EPR-Bell experiments, the atoms were excited not by
lasers but by electron bombardment \cite{Asp02}. In an accretion disc,
irradiation by a broad continuum of frequencies will excite some atoms to
appropriate states leading to cascade emission. The observed K$\alpha $ iron
line is generated by a $1s\rightarrow 2p$ vacancy transition that leaves a
vacancy in the L-shell ($n=2$). An electron in the M-shell ($n=3$) may then
fall into the L-shell (a $2p\rightarrow 3d$ vacancy transition), resulting
in the emission of a second (L$\alpha $) photon. According to the detailed
calculations of Jacobs and Rozsnyai \cite{JR86}, for an initial K-shell ($1s$%
) vacancy created in (neutral) iron, the radiative vacancy transitions $%
1s\rightarrow 2p$ and $2p\rightarrow 3d$ have respective probabilities $%
P(1s\rightarrow 2p)=0.28$ and $P(2p\rightarrow 3d)=0.18\times 10^{-2}$. (See
table I of ref. \cite{JR86}.) Because $2p\rightarrow 3d$ can occur only
after $1s\rightarrow 2p$, the probability for the cascade $1s\rightarrow
2p\rightarrow 3d$ is $P(1s\rightarrow 2p\rightarrow 3d)=P(2p\rightarrow
3d)\approx 2\times 10^{-3}$. Of an initial ensemble of iron atoms with a
K-shell ($1s$) vacancy, about $0.2\%$ will make the transition $%
1s\rightarrow 2p\rightarrow 3d$ and emit a pair of photons. And of the
subensemble that emits a K$\alpha $ photon, a fraction $P(2p\rightarrow
3d\mid 1s\rightarrow 2p)=P(2p\rightarrow 3d)/P(1s\rightarrow 2p)\approx
6\times 10^{-3}$ will subsequently emit an L$\alpha $ photon. Thus, of the K$%
\alpha $ photons that are currently observed from black-hole accretion
discs, about $0.6\%$ should be accompanied by L$\alpha $ cascade photons.

However, even if the K$\alpha $--L$\alpha $ pairs are strongly entangled
(which we have not established), $99.4\%$ of the observed K$\alpha $ photons
will have no cascade partners at all, so that even if the suggested
deviations from Malus' law exist they will be greatly diluted. Further,
while an efficient X-ray polarimeter has been developed for astrophysical
observations in the $2-10\ \mathrm{keV}$ band \cite{Costa01}, the device
(based on the photoelectric effect) unfortunately destroys the measured
photon and cannot be used for two successive polarisation measurements on
the same photon.

The proposed experiment must therefore await the detection of other
relativistically broadened lines, with a larger fraction of entangled
photons, and in a frequency band that is more convenient for the required
polarisation measurements. (Broadened lines from oxygen, nitrogen and carbon
have been reported \cite{Ogle04}.)

It is possible that the sought-for deviations from Malus' law, even if they
exist, could be smeared out by averaging over the finite size of the
emitting region. This might happen if the distribution $\rho (\lambda )\neq
\rho _{\mathrm{QT}}(\lambda )$ depends on the location of the emission, but
such averaging will have no effect at all if $\rho (\lambda )$ is
uncorrelated with location. A further possible complication is that
scattering along the line of sight could cause nonstandard distributions of
hidden variables to relax back to the quantum distribution $\rho _{\mathrm{QT%
}}(\lambda )$. This will depend on the hidden-variables model, as well as on
the degree of scattering (which again may be controlled by using an
appropriate frequency band or a face-on view). In any case, the proposed
experiment would nevertheless constitute an extreme test of quantum theory,
and a positive result would of course be of great interest.

Polarisation measurements will always show deviations from $\cos ^{2}\Theta $
due to ordinary noise and experimental errors. These may be distinguished
from genuine deviations from Malus' law by comparing the results obtained
from the astronomical source with results obtained from a comparable photon
source in the laboratory. Further, if the effect exists it will be larger
for photons closer to the red end of the extended red wing of the emission
line, as these are more likely to have partners that were captured (having
been emitted closer to the horizon).

In conclusion, it would be worthwhile to test Malus' law for single photons
in extreme conditions. Photons with entangled partners inside black holes
seem of particular interest. Such an experiment could be carried out, upon
identification of an appropriate atomic cascade close to a black-hole
horizon. It remains to be seen if such cascades will in fact be identified.

\end{document}